\documentclass[10pt,letterpaper,twocolumn]{article} 

\usepackage{ol2}
\usepackage[draft]{hyperref}
\usepackage{amsmath}

\begin{document}

\twocolumn[ 

\title{Optical parametric amplification and oscillation assisted by low-frequency stimulated emission}


\author{Stefano Longhi}

\address{Dipartimento di Fisica, Politecnico di Milano and Istituto di Fotonica e Nanotecnologie del Consiglio Nazionale delle Ricerche, Piazza L. da Vinci 32, I-20133 Milano, Italy (stefano.longhi@polimi.it)}

\begin{abstract}
 Optical parametric amplification/oscillation provide a powerful tool for coherent light generation in spectral regions inaccessible to lasers. Parametric gain is based on a frequency {\it down-conversion} process, and thus it can not be realized for signal waves at a frequency $\omega_3$ {\it higher} than the frequency of the pump wave $\omega_1$. In this work we suggest a route toward the realization of {\it up-conversion} optical parametric amplification and oscillation, i.e. amplification of the signal wave by a coherent pump wave of lower frequency, assisted by stimulated emission of the auxiliary idler wave. When the signal field is resonated in an optical cavity, parametric oscillation is obtained. Design parameters for the observation of up-conversion optical parametric oscillation at $\lambda_3=465$ nm are given for a periodically-poled lithium-niobate  (PPLN) crystal doped with Nd$^{3+}$ ions. 
\end{abstract}
 ] 

Optical frequency conversion in $\chi^{(2)}$ nonlinear crystals, such frequency doubling, sum and difference frequency generation, 
optical parametric amplification (OPA) and oscillation (OPO), provides a powerful and flexible tool for the generation and amplification of 
light waves in spectral regions inaccessible to lasers \cite{r1}. Since more than four decades, a wide variety of coherent light sources and methods based on frequency conversion processes have been realized with important applications to the generation of of twin photons and other nonclassical states of light \cite{r2,r4}, broadband coherent light generation  \cite{r5,r6}, ultrafast time-domain spectroscopy \cite{r7}, high-resolution spectroscopy and metrology \cite{r8,r10}, and optical sensing \cite{r12,r13}. In the past few years, several novel schemes of frequency conversion have been introduced, including cascading and multi-step optical parametric processes \cite{r14,r14uff}, mirror-less OPO \cite{r14bis,r14tris,r15,r15bis}, adiabatic frequency conversion \cite{r16,r17,r18}, time-reversal optical parametric oscillation \cite{r19}, hybridly-pumped OPOs \cite{r20,r20bis},  and quasi-parametric optical amplification \cite{r21}, etc.\\
In ordinary OPO/OPA devices, parametric gain is based on a frequency {\it down-conversion} process: a weak optical field at low frequency $\omega_1$ (signal wave) is amplified  by a strong and of higher frequency $\omega_3$ field (pump wave) under phase matching condition. A pump photon in the crystal annihilates and creates a down-converted pair of photons, one at frequency $\omega_1$ (thus providing signal amplification) and the other one at frequency $\omega_2= \omega_3-\omega_1$ (the idler wave). When optical feedback (i.e. a cavity) is provided for the signal wave, parametric oscillation is obtained [Fig.1 (a)]. However, parametric gain can not be provided for coherent waves at frequencies higher than the pump wave. Such a possibility would be extremely useful to generate and amplify high-frequency coherent waves (in the uv and even X-ray spectral regions), without recurring to low-efficienty schemes like e.g. higher-order parametric processes \cite{uff}.\\ In this Letter a route toward the realization of OPA and OPO devices, with a pump wave at frequency $\omega_1$ smaller than the signal frequency $\omega_3$, is theoretically suggested. The main idea, sketched in Figs.1(b) and (c), is to exploit optical gain by stimulated emission available at the lower frequency $\omega_2$ of the idler wave, and to transfer such a gain into the higher-frequency signal field by nonlinear frequency conversion. Can we realize optical gain for the signal wave at the high frequency $\omega_3$, which can not be provided by stimulated emission? What we can do, is to down convert the signal wave to a low-frequency field $\omega_2$ (the idler wave), for which optical gain by stimulated emission in an incoherently pumped active  medium is available. After amplification, frequency up-conversion, from $\omega_2$ to $\omega_3$, is achieved, which results in an overall amplification of the original signal wave at frequency $\omega_3$. 
A schematic of the three-step parametric amplification process assisted by stimulated emission is shown in Fig.1(c). For the sake of simplicity, we consider a single nonlinear crystal with quasi phase matching (QPM) obtained by domain inversion and pumped by a CW wave at frequency $\omega_1$. The crystal of length $L=2l+l_g$ comprises three sections: a first QPM section of length $l$ that yields  perfect phase matching ($\sigma=1$ region) for the $\omega_2=\omega_3-\omega_1$ interaction and generates the idler wave via dow-conversion; a second section of length $l_g$ without the QPM grating ($\sigma=0$ region) and doped with an active medium that provides optical gain for the idler wave; and a third  QPM section of length $l$ with reversed sign of the effective nonlinearity ($\sigma=-1$ region) that realizes back conversion. The equations describing nonlinear interaction of the three fields in the QPM crystal with assisted gain at the idler frequency read \cite{r22,r23}  
\begin{figure}[htb]
\centerline{\includegraphics[width=8.4cm]{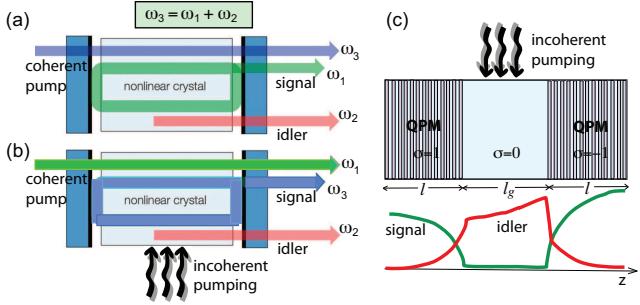}} \caption{ \small
(Color online)  (a) Schematic of an ordinary OPO device in a singly-resonant configuration. (b) Schematic of an up-conversion OPO assisted by stimulated emission of the idler wave. As compared to an ordinary OPO, the frequencies of signal and pump waves are interchanged, and optical gain by stimulated emission for the idler wave is provided by incoherent pumping of the crystal. (c) Schematic of the nonlinear crystal used in the up-conversion OPO. The crystal comprises three sections: two QPM grating sections of same length $l$, separated by a gain region of length $l_g$ for the idler wave.The idler gain is realized by stimulated emission, with population inversion in the doped crystal provided by some incoherent pumping. The lower panel in (c) shows schematically the evolution of the signal and idler intensities along the nonlinear crystal: down-conversion ($\omega_3 \rightarrow \omega_1+ \omega_2$) in the first QPM section of the crystal, amplification of the idler wave in the intermediate section by stimulated emission, and up-conversion ($\omega_2 + \omega_1 \rightarrow \omega_3$)  in the last  QPM crystal section.}
\end{figure}
\begin{subequations}
\begin{eqnarray}
\frac{\partial B_1}{\partial z}+\frac{1}{v_{g1}} \frac{\partial B_1}{\partial t} & = & i [\sigma(z)/l] B_2^* B_3 \\
\frac{\partial B_2}{\partial z}+\frac{1}{v_{g2}} \frac{\partial B_2}{\partial t} & = & i [\sigma(z)/l] B_1^* B_3+g(z)B_2 \\
\frac{\partial B_3}{\partial z}+\frac{1}{v_{g3}} \frac{\partial B_3}{\partial t} & = & i [\sigma(z)/l] B_1 B_2 
\end{eqnarray}
\end{subequations}
In the above equations, $B_m$ ($m=1,2,3$) are dimensionless slowly-varying envelopes of the pump, idler and signal fields at frequencies $\omega_1$, $\omega_2$, and $\omega_3=\omega_1+\omega_2$, respectively; $v_{gm}$ are the corresponding group velocities; $\sigma(z)$ is the normalized strength of the nonlinear interaction in the QPM crystal ($\sigma=1$ for  $0<z<l$; $\sigma=0$ for $l<z<l_g+l$; $\sigma=-1$ for $l+l_g<z<L$); and $g(z)$ is stimulated-emission gain for the idler wave [$g(z)=0$ for $0<z<l$, $l+l_g<z<L$ and $g(z)=g_0$ for $l<z<l+l_g$]. The intensities $I_m$ of the three waves are given by
 $I_m=\epsilon_0 c_0 n_1n_2n_3 \lambda_1 \lambda_2 \lambda_3 |B_m|^2 /( 8 \pi^2 l^2 \lambda_m d_{eff}^2)$ ($m=1,2,3$), where $\lambda_m= 2 \pi c_0/ \omega_m$ are the wavelengths in 
vacuum of the three waves, $n_m$ the corresponding refractive indices, and $d_{eff}$ is the effective nonlinear $d$-coefficient in the QPM crystal. For a $+/-$ square-wave first-order QPM grating with $50\%$ duty cycle and period $\Lambda=|n_3/ \lambda_3-n_2 / \lambda_2 - n_1 / \lambda_1|^{-1} $, one has $d_{eff}=(2/ \pi)d$, where $d=(1/2) \chi^{(2)}$ is the second-order nonlinear coefficient of the bulk crystal. The sign change of the effective nonlinearity ($\sigma=\pm 1$) between the first and last sections of the crystal can be obtained, for example, by a $\pi$ phase slip of the QPM gratings in the two sections \cite{r23}. The parametric gain for the high-frequency field $\omega_3$  is obtained by looking for a solution to Eqs.(1) of the form $B_{2,3}(z,t)=A_{2,3}(z) \exp(i \Omega t)$ and $B_1(z,t)=A_1(z)$ with the boundary condition $A_2(0)=0$, where $\Omega$ is the frequency offset from $\omega_3$ and $G(\Omega)=|A_3(L)/A_3(0)|^2$ is the spectral power gain coefficient. The parametric gain turns to depend on the effective frequency detuning parameter $\delta l =(1/v_{g2}-1/v_{g3})l \Omega /2$, the dimensionless intensity $\mathcal{I}_1=|B_1(0)|^2$ of the incident pump beam, the gain amplification $g_0l_g$ of the idler wave, and the ratio $x=l_g/l$. In the undepleted pump approximation, i.e. for $|B_3(0)| \ll |B_1(0)|$,  the (unsaturated) gain is independent of the intensity level $\mathcal{I}_3=|B_3(0)|^2$ of the incident signal wave and reads explicitly
\begin{equation}
G(\delta l)  =  \left| \frac{\mathcal{I}_1}{\rho^2} \exp(-2 i \delta l x+g_0l_g) \sin^2 \rho+ \left[ \cos \rho+i (\delta l / \rho) \sin \rho \right]^2 \right|^2
\end{equation}
\begin{figure}[htb]
\centerline{\includegraphics[width=8.7cm]{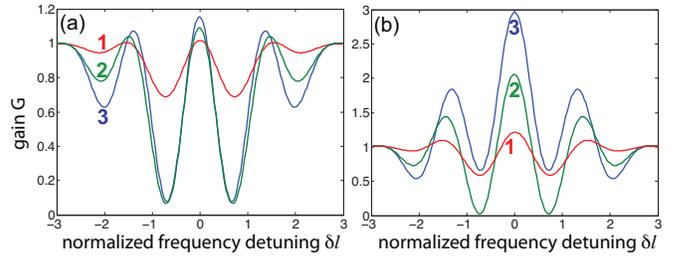}} \caption{ \small
(Color online)  Behavior of the power signal gain $G$ versus normalized frequency detuning $\delta l$ in the undepleted pump regime for increasing values of the normalized pump intensity $\mathcal{I}_1$. Curve 1: $\mathcal{I}_1=0.1$; curve 2:  $\mathcal{I}_1=0.5$;  curve 3: $\mathcal{I}_1=1$. In (a) $g_0l_g=0.1$; in (b) $g_0l_g=0.7$. }
\end{figure}
\begin{figure}[htb]
\centerline{\includegraphics[width=8.7cm]{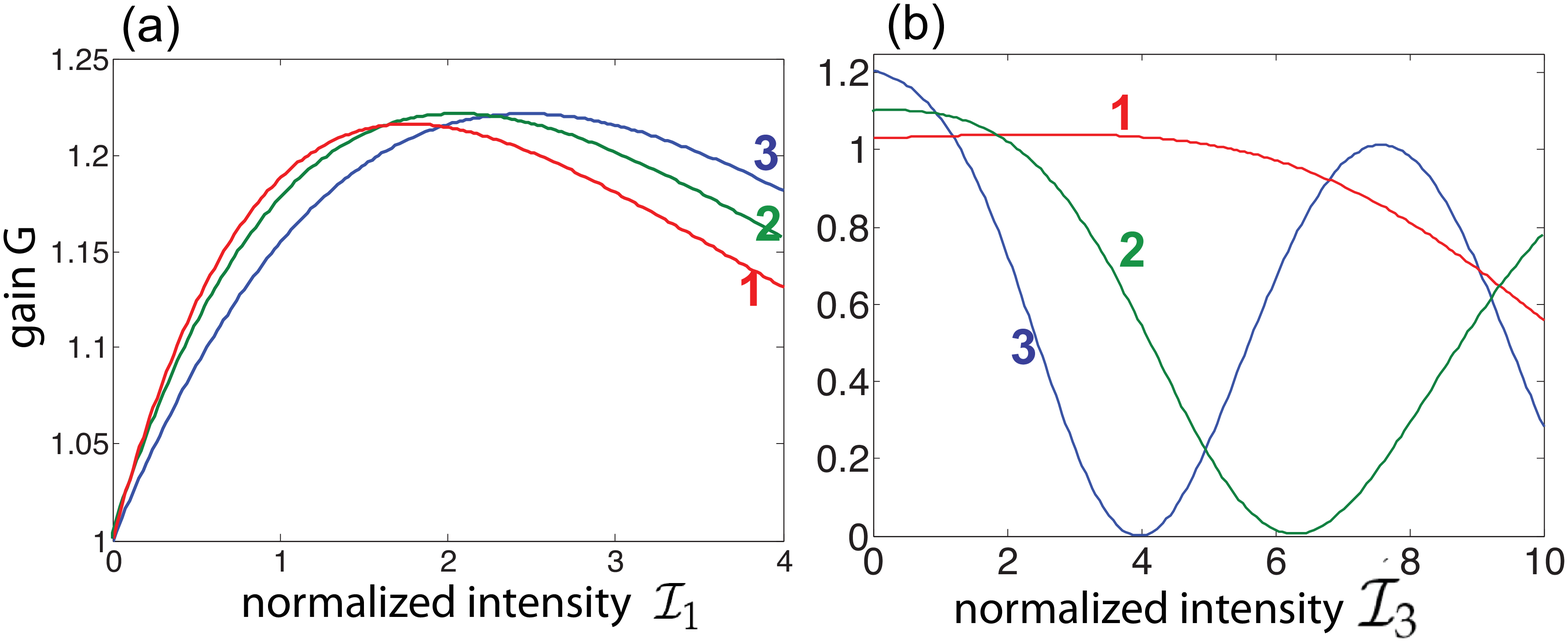}} \caption{ \small
(Color online) Numerically-computed behavior of the signal gain $G$ in the depleted pump regime at resonance $\delta l=0$. (a) Gain $G$ versus the normalized pump intensity $\mathcal{I}_1$ for $g_0l_g=0.1$, $x=1$, and for a few values of the input signal intensity $\mathcal{I}_3$. Curve 1: $\mathcal{I}_3=2$; curve 2: $\mathcal{I}_3=1$; curve 3: $\mathcal{I}_3 \rightarrow 0$ (undepleted pump limit). (b) Gain $G$ versus the normalized intensity $\mathcal{I}_3$ of incident signal wave for $\mathcal{I}_1=0.1$,  $x=1$, and for a few values of the gain parameter $g_0l_g$. Curve 1: $g_0l_g=0.1$; curve 2: $g_0l_g=0.4$; curve 3: $g_0l_g=0.7$.}
\end{figure}

where we have set $\rho \equiv \sqrt{(\delta l)^2 + \mathcal{I}_1}$. Figure 2 shows typical spectral gain curves in the undepleted pump regime for increasing values of the pump intensity $\mathcal{I}_1$ and for two different values of the gain parameter $g_0l_g$ and for $l_g/l=1$. Note that the maximum gain is obtained at $\delta l=0$, i.e. under perfect phase matching in the nonlinear crystal, and reads $G_{max}=[1+ (\exp(g_0l_g)-1) \sin^2 \sqrt{\mathcal{I}_1}]^2$. As a function of the pump intensity $\mathcal{I}_1$,  $G_{max}$ shows an oscillatory behavior, with maxima equal to $\exp(2 g_0 l_g)$ at intensity levels such that $ \sqrt{\mathcal{I}_1}$ is an integer multiple than $(\pi /2)$. Such a behavior is readily explained after observing that, at $\mathcal{I}_1= \pi^2/4$, in the first section of the crystal the signal wave at frequency $\omega_3$ is completely converted into the idler wave, which is then amplified by the power factor $\exp(2g_0 l_g)$ in the middle crystal section and finally fully up-converted into the signal field at frequency $\omega_3$ in the last section of the crystal. For a pump intensity $\mathcal{I}_1$ smaller than $\pi^2/4$, the signal/idler conversion is incomplete, resulting in a lowered gain $G$ for the signal wave.  

 In the strong-signal regime, the spectral gain also depends on the incident signal intensity $\mathcal{I}_3=|B_3(0)|^2$ and the analytical expression of the spectral gain $G$ is rather cumbersome, since the exact solution to Eqs.(1) involves elliptic integrals \cite{r22}. However, in this regime the spectral gain curve can be readily obtained  by direct numerical integration of Eqs.(1). As in the small-signal regime, the maximum gain is attained at exact phase matching $\delta l =0$. Figure 3 shows the numerically-computed behavior of the power gain $G$ at $\delta l=0$ as a function of the pump intensity $\mathcal{I}_1$ [panel (a)] and signal intensity $\mathcal{I}_3$ [panel (b)], and for a few values of the gain parameter $g_0l_g$. Note that, for small values of the gain parameter and pump power [curve 1 in Fig.3(b)], the power gain $G$ at the signal frequency $\omega_3$ shows first an {\it increase} (rather than a decrease) as $\mathcal{I}_3$ is increased from zero, indicating that there is not gain saturation in this regime.\\
 Given the optical gain at the signal frequency $\omega_3$ provided by the nonlinear crystal of Fig.1(c), an OPO can be realized by resonating the signal field in an optical cavity [Fig.1(b)]. We remark that such an "up-conversion" OPO is conceptually very different than previously demonstrated hybridly-pumped singly-resonant OPOs \cite{r20,r20bis}, where stimulated-emission gain is provided for the signal wave to decrease OPO threshold, or intracavity frequency up-conversion schemes \cite{r23bis}, that need an externally-injected field for wave mixing. With reference to Fig.1(b), let us assume that the two resonator mirrors are fully transparent at the pump and idler wavelengths, whereas they show reflectances $R_1=1$ and $R_2=R<1$ at the signal wavelength. Since the maximum parametric gain is attained at exact phase matching $\delta l=0$, the cavity length is tuned so that a cavity axial mode is in resonance with the signal wave at frequency $\omega_3$. In the OPO scheme, there is not any seeding signal and oscillation is started from quantum noise at the signal frequency, like in an ordinary singly-resonant OPO. Neglecting saturation of gain $g_0$ for the idler wave and indicating by $B_3^{(n)}$ the amplitude of the intracavity signal field at the $n$-th round trip in the resonator, the dynamical evolution of $B_3^{(n)}$ at successive transits in the cavity is described by the nonlinear map 
 \begin{equation}
 B_3^{(n+1)} =\sqrt{R} \mathcal {P} (B_3^{(n)})
 \end{equation}
 where $\mathcal{P}(B_3^{(n)})$ is a nonlinear operator of $B_3^{(n)}$ which describes the propagation of the field $B_3(z)$ in the crystal [Eqs.(1)], from $z=0$ to $z=L$, subjected to the boundary conditions $B_3(0)=B_3^{(n)}$, $B_2(0)=0$, $B_1(0)= \sqrt{\mathcal{I}_1}$. The expression of $B_3(L)=\mathcal{P}(B_3^{(n)})$ is obtained from the exact solution to Eqs.(1) in each crystal section, with $dB_l/dt=0$, and involves combinations of sn, cn and dn Jacobi elliptic functions \cite{r22}. For a given value of the gain parameter $g_0l_g$, the input-output curve $\mathcal{I}_3=\mathcal{I}_3 ( \mathcal{I}_1)$ of the OPO is obtained by looking for the fixed points of the map (3). Once the fixed points have been determined, their stability can be checked by perturbing the initial condition of the map (3) and propagating at successive transits in the cavity. The switch-on dynamics of the OPO is obtained by iteration of the map (3) starting from a small amplitude of $B_3^{(n=1)}$ at the first round-trip, which mimics semi-classically the quantum noise at the signal frequency in the cavity. Figure 4 shows two typical examples of input-output OPO curves corresponding to two different values of the linear gain parameter $g_0l$. Solid and dotted curves correspond to stable and unstable fixed points of the map (3), respectively. Note that, for a small gain parameter $g_0l_g$, the OPO threshold shows a subcritical behavior with a bistability loop (hysteresis) close to the threshold pump value [arrows in Fig.4(a)], whereas the OPO threshold is supercritical for larger values of the gain [Fig.4(b)]. The subcritical behavior of the OPO threshold in the former case can be explained after observing that the parametric gain of the signal wave in the nonlinear crystal versus $\mathcal{I}_3$, at small values of the gain parameter $g_0l_g$, shows a first increase with $\mathcal{I}_3$, while gain saturation occurs at large values of $\mathcal{I}_3$ [curve 1 in Fig.3(b)].  
  \begin{figure}[htb]
\centerline{\includegraphics[width=8.7cm]{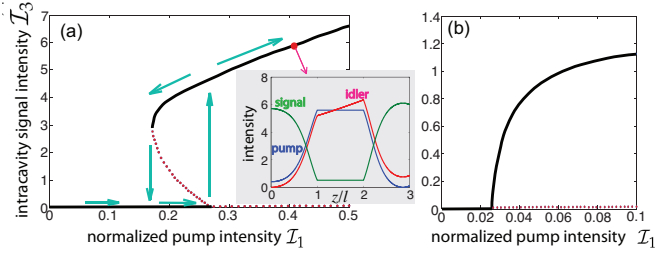}} \caption{ \small
(Color online) Numerically-computed OPO input-output curve for parameter values $x=l_g/l=1$, $R= 95 \%$ (output coupler), and for (a) $g_0l_g=0.1$, (b)  $g_0l_g=0.7$. The intracavity intensity $\mathcal{I}_3$ of the steady-state signal field refers to the wave just after reflection from the output mirror. The output signal intensity is given by $(1-R) \mathcal{I}_3/R$. Solid and dotted curves correspond to stable and unstable fixed points of the map (3). The inset in (a) shows, as an example, the distribution of normalized intracavity intensities for pump, signal and idler waves along the nonlinear crystal at steady-state for an incident pump intensity $\mathcal{I}_1=0.4$. For an initial noise level $B_3^{(n=1)}=10^{-5}$ of signal field in the cavity, the switch-on time of the OPO is about $n=1000$ round trips.}
\end{figure}

In a slightly modified up-conversion OPO scheme, the coherent pump beam at frequency $\omega_1$ is used to provide population inversion in the middle section of the crystal as well. With respect to the scheme of Figs.1(b) and (c), such a configuration offers the advantage of requiring a single pump beam. However, in this case the pump wavelength, and thus the OPO signal wavelength,  is constrained by the absorption properties of the active medium. The theoretical analysis of the singly-pumped up-conversion OPO device should be modified to account for the linear absorption of the pump wave  at $\lambda_1$ in the middle section of the crystal, and the intensity-dependent behavior of both absorption coefficient $\alpha$ at $\lambda_1$ and of the gain coefficient $g$ at $\lambda_2$. If the stimulated emission  at the idler wavelength is achieved by doping the nonlinear crystal with a four-level amplifier system (such as Nd$^{3+}$) with density $N_t$, from simple rate equation analysis one can derive the following expressions for absorption and gain coefficients per unit length in the region $l<z<l+l_g$
\begin{eqnarray}
2 \alpha (z) & = &  \frac{\sigma_a N_t}{1+ \eta (I_1/I_a)(1+I_2/I_s)^{-1}} \\
2 g(z) & = & \sigma_e N_t \eta (I_1/I_a) \times \frac{1}{1+I_2/I_s+\eta I_1/I_a},
\end{eqnarray}
 where: $I_{1,2}(z)$ are the space-dependent intensities of pump and idler waves, $\sigma_a$ is the absorption cross section of the pump wave, $\sigma_e$ is the stimulated emission cross section at the idler wavelength $\lambda_2$, $\tau$ is the lifetime of the upper-level gain transition, $\eta < 1$ is the fraction of the absorbed pump photons that contribute to the population inversion of the gain transition, $I_s= \hbar \omega_2/(\sigma_e \tau)$ is the saturation intensity of the gain transition, and $I_a= \hbar \omega_1/(\sigma_a \tau)$. An example of an input-output curve for a singly-pumped up-conversion OPO device, which accounts for space-dependence and saturation effects of absorption and gain coefficients, is shown in Fig.5 for parameter values that apply to Nd$^{3+}$-doped LiNbO$_3$ OPO pumped at $\sim 800$ nm.

\begin{figure}[htb]
\centerline{\includegraphics[width=8cm]{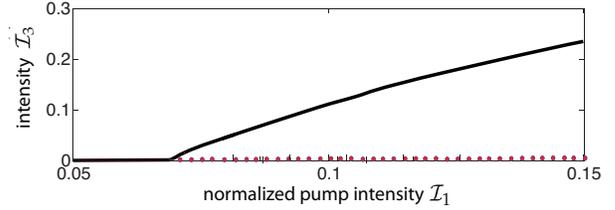}} \caption{ \small
(Color online) Numerically-computed input-output curve of a singly-pumped up-conversion OPO for parameter values $x=l_g/l=1$, $R= 90 \%$ with 
intensity-dependent absorption and gain coefficients at $\lambda_1$ and $\lambda_2$ described by the relations $2 \alpha l= 0.55/ \{1+ 15.386 \times \mathcal{I}_1(z)[1+48.2 \times \mathcal{I}_2(z)]^{-1} \}$ and $2 g l = 26.157  \times  \mathcal{I}_1(z)  /[1+ 48.2 \times \mathcal{I}_2(z)+15.386 \times \mathcal{I}_1(z)]$, respectively. Such relations apply to Nd$^{3+}$-doped LiNbO$_3$ crystals for parameter values given in the text.}
\end{figure}

 To check the feasibility of the up-conversion OPO scheme proposed in the present work, let us consider as an example parametric oscillation of  signal wave in the blue region at $\lambda_3=465$ nm using a PPLN crystal doped with Nd$^{3+}$ and pumped at $\lambda_1=814$ nm. Lithum Niobate (LNB), besides of providing a high nonlinear $d$ coefficient ($d=d_{33} \simeq 27$ pm/V), is a suitable matrix host for rare-earth solid-state lasers in the near-infrared \cite{r25,r26,r27,r28}. For example, gain coefficients as high as $g_0 \sim 6.35$ dB/cm at $\lambda_2=1085$ nm have been demonstrated in waveguide LNB crystals doped with Nd$^{3+}$ ions and optically-pumped at around 800 nm, with 22 mW absorbed pump power in a 5.9-mm long crystal \cite{r26}. To realized the up-conversion OPO, let us consider a PPLN crystal of length $L=3$ cm, with $l=l_g=1$ cm, pumped by a coherent CW wave at wavelength $\lambda_1=814$ nm, which can be provided by a GaAs/AlGaAs semiconductor laser. Note that the same wavelength is used to optically pump the Nd$^{3+}$ ions in the LNB crystal. First-order QPM at such wavelengths is attained by a grating period $\Lambda \simeq 4.457 \; \mu$m, taking into account that $n_m=2.155, 2.1744, 2.2704$ and $v_{gm} / c_0=0.4639, 0.4598, 0.4404$ ($m=1,2,3$), as obtained from Sellmeier equations. The intensities in physical units are given $I_1 \simeq 0.6108 \times \mathcal{I}_1 \; {\rm MW/cm^2}$, $I_2 \simeq 0.458  \times \mathcal{I}_2 \; {\rm MW/cm^2}$, and $I_3 \simeq 1.07 \times  \mathcal{I}_3 \; {\rm MW/cm^2}$. Assuming typical parameter values $\sigma_a=5.5 \times 10^{-20} \; {\rm cm}^2$, $\sigma_e=1.7 \times 10^{-19} \; {\rm cm}^2$, $\tau=114 \; \mu$s, and $\eta=0.98$ \cite{r28}, for a Nd$^{3+}$ doping of $N_t=2 \times 10^{19} \; {\rm ions/cm^3}$, one has $2 \alpha l= 0.55/ \{1+ 15.386 \times \mathcal{I}_1(z)[1+48.2 \times \mathcal{I}_2(z)]^{-1} \}$ and $2 g l = 26.157  \times  \mathcal{I}_1(z)  /[1+ 48.2 \times \mathcal{I}_2(z)+15.386 \times \mathcal{I}_1(z)]$.  For an output coupling $T=1-R=10\%$ at $\lambda_3$, the OPO input-output curve is the one shown in Fig.5. In physical units, the OPO pump threshold is given by $I_{1th} \simeq 41 \; \rm{kW/cm^2}$, which for an
 effective mode are $A_e \sim 10 \; \mu m^2$ corresponds to the relatively low pump power threshold $P_{1th} \simeq A_e I_{1 th} \simeq 4.1$ mW. For a pump power twice its threshold value, an output power $P_3 \simeq  A_e R I_3/(R-1) \simeq 2.3$ mW in the blue is estimated from the curve of Fig.5 and assuming approximately the same area for the signal and pump waveguide modes. 
 
 In conclusion, we have suggested the possibility to realize parametric amplification and oscillation of a high-frequency signal wave coherently pumped by a lower frequency pump wave, assisted by stimulated emission of the auxiliary idler wave. The feasibility of the proposed scheme has been discussed for a PPLN-based compact OPO waveguide doped with Nd$^{3+}$ ions, emitting in the blue. The "up-conversion" OPO provides a conceptually novel method of nonlinear frequency manipulation that might be useful in the design of compact light sources at short wavelengths.

\newpage


 {\bf References with full titles}\\

1. R.W. Boyd, {\it Nonlinear Optics} (3rd edition, Academic, New York, 2008).\\ 
2. L. Wu, H. J. Kimble, J. Hall and H. Wu, {\it Generation of Squeezed states of light by Parametric Down conversion}, Phys. Rev. Lett. {\bf 57}, 2520 (1986).\\      
3. J. Mertz, T. Debuisschert, A. Heidmann, C. Fabre, and E. Giacobino, {\it Improvements in the observed intensity correlation of optical parametric oscillator twin beams}, Opt. Lett. {\bf 16}, 1234 (1991).\\
4. G. Cerullo and S. De Silvestri, {\it Ultrafast optical parametric amplifiers}, Rev. Sci. Instrum. {\bf 74}, 1 (2003).\\
5. M. Ebrahim-Zadeh, S. C. Kumar, and K. Devi, {\it Yb-Fiber-Laser-Pumped Continuous-Wave Frequency Conversion Sources from the Mid-Infrared to the Ultraviolet}, IEEE J. Selc. Top. Quantum Electron. {\bf 20}, 0902823  (2014).\\
6. C.Manzoni, D. Polli, and G. Cerullo, {\it Two-color pump-probe system broadly tunable over the visible and the near infrared with sub-30 fs temporal resolution}, Rev. Sci. Instrum. {\bf 77}, 3103 (2006).\\ 
7. P. H. S. Ribeiro, C. Schwob, A. Maitre, and C. Fabre, {\it Sub-shot-noise high-sensitivity spectroscopy with optical parametric oscillator twin beams}, Opt. Lett. {\bf 22}, 1893 (1997).\\
8. J. Krieg, A. Klemann, I. Gottbeh\"{u}t, S. Thorwirth, T. F. Giesen, and S. Schlemmer, {\it A continuous-wave optical parametric oscillator around 5-$\mu$m wavelength
for high-resolution spectroscopy}, Rev. Sci. Instrum. {\bf 82},  063105 (2011).\\
9. K. Fradkin-Kashi, A. Arie, P. Urenski, and G. Rosenman, {\it Mid-infrared difference frequency generation in
periodically poled KTiOAsO$_4$ and application to gas sensing}, Opt Lett {\bf 25}, 743 (2000).\\
10. J. Peng, {\it Developments of mid-infrared optical parametric oscillators for spectroscopic sensing: a review}, Opt. Eng. {\bf 53}, 061613 (2014).\\
11. S.M. Saltiel, A.A. Sukhorukov, and Y.S. Kivshar, {\it Multistep Parametric Processes in Nonlinear Optics}, Prog. Opt. {\bf 47}, Ed. E. Wolf (Elsevier, Amsterdam, 2005), pp. 1-73.\\
12. M. Levenius, M. Conforti, F. Baronio, V. Pasiskevicius, F. Laurell, C. De Angelis, and K. Gallo, {\it Multistep quadratic cascading in broadband optical parametric generation}, Opt. Lett. {\bf 37}, 1727 (2012).\\
13. Y.I. Ding and J.B. Khurgin, {\it Backward optical parametric oscillators and amplifiers}, IEEE J. Quantum Electron. {\bf 32},1574 (1996).\\
14. Y. J. Ding, S. J. Lee, and J. B. Khurgin, {\it Transversely Pumped Counterpropagating Optical Parametric Oscillation and Amplification}, Phys. Rev. Lett. {\bf 75}, 429 (1995).\\
15. C. Canalias and V. Pasiskevicius, {\it Mirrorless optical parametric oscillator}, Nature Photon. {\bf 1}, 459 (2007).\\
16. C. Montes, P. Aschieri, and M. de Micheli, {\it Backward optical parametric efficiency in quasi-phase-matched GaN waveguide presenting stitching faults}, Opt. Lett. {\bf 38}, 2083 (2013).\\
17. H. Suchowski, V. Prabhudesai, D. Oron, A. Arie, and Y. Silberberg, {\it Robust adiabatic sum frequency conversion }, Opt. Express {\bf 17}, 12731 (2009).\\
18. J. Moses, H. Suchowski, and F.X. K\"{a}rtner, {\it Octave-spanning coherent mid-IR generation via adiabatic difference frequency conversion}, Opt. Lett. {\bf 37}, 1589 (2012).\\
19. H. Suchowski, G. Porat, and A. Arie, {\it Adiabatic processes in frequency conversion}, Laser \& Photon. Rev. {\bf 8}, 333 (2014).\\
20. S. Longhi, {\it Time-Reversed Optical Parametric Oscillation}, Phys. Rev. Lett. {\bf 107}, 033901 (2011).\\
21. I. Breunig, J. Kiessling, B. Knabe, R. Sowade, and K. Buse, {\it Hybridly-pumped continuous-wave optical parametric oscillator}, Opt. Express {\bf 16}, 5662 (2008).\\
22. M. Siltanen, T. Leinonen, and L. Halonen, {\it Decreased oscillation threshold of a continuous-wave OPO using a semiconductor gain mirror}, Opt. Express {\bf 19}, 19675 (2011).\\
23. J. Ma, J. Wang, P. Yuan, G. Xie, K. Xiong, Y. Tu, X. Tu, E. Shi, Y. Zheng, and L. Qian,  
{\it Quasi-parametric amplification of chirped pulses based on a Sm$^{3+}$-doped yttrium calcium oxyborate crystal}, Optica {\bf 2}, 1006 (2015).\\
24. S. Meyer, B. N. Chichkov, and B. Wellegehausen, {\it High-order parametric amplifiers}, J. Opt. Soc. Am. B {\bf 16}, 1587 (1999).\\
25. R. A. Baumgartner and R. L. Byer, {\it Optical parametric amplification}, IEEE J. Quantum Electron. {\bf 15}, 432 (1979).\\
26. S. Longhi, M. Marano, and P. Laporta, {\it Dispersive properties of quasi-phase-matched optical parametric amplifiers}, Phys. Rev. A {\bf 66}, 033803 (2002).\\
27. S. Shichijyo, K. Yamada, and K. Muro, {\it Efficient intracavity sum-frequency generation of 490-nm radiation by use of potassium niobate}, Opt. Lett. {\bf 19}, 1022 (1994).\\
28. A. Cordova-Plaza, T. Y. Fan, M. J. F. Digonnet, R. L. Byer, and H. J. Shaw, {\it Nd:MgO:LiNbO$_3$ continuous-wave laser pumped by a laser diode}, Opt. Lett. {\bf 13}, 209 (1988).\\
29. E. Lallier, J. P. Pocholle, M. Papuchon, M. P. De Micheli, M. J. Li, Q. He, D. B. Ostrowsky, C. Grezes-Besset, and E. Pelletier, {\it Nd:MgO: LiNbO$_3$ Channel Waveguide Laser Devices}, IEEE J. Quantum Electron. {\bf 27}, 618 (1991).\\ 
30. R. Brinkmann, W. Sohler, and H. Suche, {\it Continuous-Wave Erbium-Diffused LiNbO$_3$ Waveguide-Laser}, Electron. Lett. {\bf 27}, 415 (1991).\\
31. R. E. Di Paolo, E. Cantelar, P.L. Pernas, G. Lifante, and F. Cusso, {\it Continuous wave waveguide laser at room temperature in Nd$^{3+}$-doped Zn:LiNbO$_3$}, Appl. Phys. Lett. {\bf 79}, 4088 (2001).

\end{document}